# Elements of a dielectric laser accelerator

J. McNeur[1,*], M. Kozák[1], N. Schönenberger[1], K. J. Leedle[2], H. Deng[2], A. Ceballos[2], H. Hoogland[3], A. Ruehl[4], I. Hartl[4], R. Holzwarth[3], O. Solgaard[2], J.S. Harris[2,5], R.L. Byer[5], P. Hommelhoff[1]

[1] Department of Physics, Friedrich-Alexander-Universität Erlangen-Nürnberg (FAU), Staudtstrasse 1, 91058 Erlangen, Germany, EU
[2] Department of Electrical Engineering, Stanford University, Stanford, California 94305, USA
[3] Menlo Systems GmbH, Am Klopferspitz 19a, 82152 Martinsried, Germany, EU
[4] Deutsches Elektronen-Synchrotron DESY, D-22607 Hamburg, Germany, EU
[5] Department of Applied Physics, Stanford University, Stanford, California 94305, USA
[*] e-mail: joshua.mcneur@fau.de

**The widespread use of high energy particle beams in basic research[1-3], medicine[4,5] and coherent X-ray generation[6] coupled with the large size of modern radio frequency (RF) accelerator devices and facilities has motivated a strong need for alternative accelerators operating in regimes outside of RF. Working at optical frequencies, dielectric laser accelerators (DLAs) – transparent laser-driven nanoscale dielectric structures whose near fields can synchronously accelerate charged particles – have demonstrated high-gradient acceleration with a variety of laser wavelengths, materials, and electron beam parameters[7-11], potentially enabling miniaturized accelerators and table-top coherent x-ray sources[9,12]. To realize a useful (i.e. scalable) DLA, crucial developments have remained: concatenation of components including sustained phase synchronicity to reach arbitrary final energies as well as deflection and focusing elements to keep the beam well collimated along the design axis. Here, all of these elements are demonstrated with a subrelativistic electron beam. In particular, by creating two interaction regions via illumination of a nanograting with two spatio-temporally separated pulsed laser beams, we demonstrate a phase-controlled doubling of electron energy gain from 0.7 to 1.4 keV (2.5% to 5% of the initial beam energy) and through use of a chirped grating geometry, we overcome the dephasing limit of 25 keV electrons, increasing their energy gains to a laser power limited 10% of their initial energy.  Further, optically-driven transverse focusing of the electron beam with focal lengths below 200 μm is achieved via a parabolic grating geometry. These results lay the cornerstone for future miniaturized phase synchronous vacuum-structure-based accelerators.**

DLAs are enticing insofar as they can provide high energy particle beams using the well-established principle of phase-synchronous acceleration in vacuum[1,2,13,14], but with a smaller footprint, higher acceleration gradient and beam properties distinct from those available via microwave acceleration[9]. In dielectric laser acceleration, electrons traverse nanostructured dielectric geometries, gaining energy via interaction with laser-induced accelerating fields[9,15-19]. These fields are generated by imprinting a periodic spatial modulation to the perpendicularly incident laser wavefront that matches the periodicity of the structure and leads to optical near-field modes travelling along the structure surface (see Figure 1). Electrons with a velocity matching the phase velocity of one of the surface modes are accelerated if injected at an appropriate phase. Notably, DLAs are based on a vacuum scheme, similar to RF accelerators, and the imparted energy gain scales linearly with the incident optical field strength, presenting clear advantages over nonlinear acceleration schemes requiring matter[20-21].

Due to the linear interaction of the laser-induced fields with the accelerated electrons, the imparted energy gain can be extended by adding sequential interaction regions. However, all previous DLA experiments have been performed with a single structure illuminated by a single perpendicularly incident laser beam. The field of DLAs thus stands in a similar position as did the field of microwave accelerators in 1947 when acceleration of electrons in the single-cell Mark I was achieved, but many further challenges, particularly beam stability over multiple cells, remained[1]. Here we show the application of two subsequent interaction stages and their phase-dependent acceleration as a first demonstration of staging in a DLA, crucial to the development of any accelerator[20-22]. Further, we show the efficacy of a chirped DLA geometry in which the excited accelerating mode has a phase velocity that increases as nonrelativistic electrons are accelerated and report on a DLA-based focusing scheme capable of keeping electron bunches well-collimated over multiple stages of travel. Lastly, we note that the demonstration of these critical DLA elements can be straight-forwardly applied to relativistic electron beams.

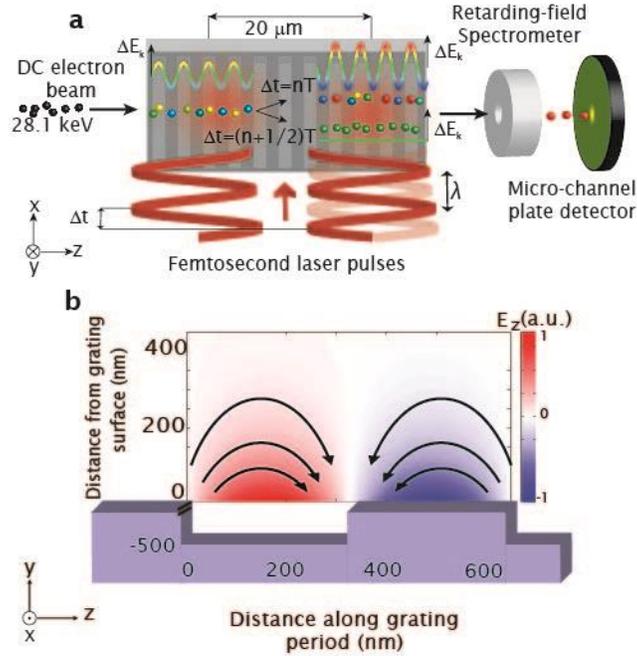

**Figure 1. Experimental setup used for demonstration of two-stage acceleration. a.** A DC electron beam (electrons indicated by spheres) interacts with two consecutive laser pulses (red curves). The acceleration of electrons during interaction with the first laser pulse (visualized by color scale) can be either doubled (when $\Delta t=nT$ for $T$ the optical cycle of the laser) or suppressed ($\Delta t=(n+1/2)T$) by the second pulse. The accelerated electrons (red) traverse a spectrometer and are detected by a micro-channel plate detector[24]. Only electrons with energy gains $\Delta E_k>(eU_s-E_{k0})$ are detected, where $e$ is electron charge and $U_s$ is the DC voltage applied to the spectrometer. **b.** The longitudinal fields of the accelerating mode, where red indicates accelerating fields and blue decelerating fields. Arrows indicate the electromagnetic forces for positively charged particles. Note that acceleration, deceleration, and deflection can occur depending on the electron injection phase.

Two-stage acceleration is demonstrated with the experimental setup shown in Figure 1. A femtosecond fiber laser operating at $\lambda \cong 2$ μm wavelength provides pulses, spatio-temporally separated when incident on a nanostructured silicon grating, with durations of $\tau \cong 500$ fs. A DC electron beam with an initial energy of $E_{k0}=28.1$ keV passes close to the surface of the grating. Depending on the relative phase of the electrons and laser fields, electrons may be either accelerated, decelerated or deflected. Therefore, to obtain optimal acceleration so that the accelerating field strength difference between the two pulses is less than 5%, the relative phase difference of the two pulses must be controlled with precision finer than $\pi/10 \approx 300$ as, achieved here with an interferometric setup.

In Figure 2a we show the dependence of the accelerated electron current on the relative phase $\Delta\varphi=2\pi\Delta t/T$ of the two pulsed laser beams measured in the temporal window $\Delta t=175-205$ fs (the electron travel time between the centers of the two interaction regions is $t_{tr}=190\pm10$ fs), with $T=6.5$ fs the optical period and $\Delta t$ the temporal difference in arrival time of the two beams. The current of electrons with energy gains higher than 400 eV is displayed along with numerical results. Evidently, for a phase difference of $\Delta\varphi=\pm 2\pi n$, net acceleration is most effective in the two acceleration regions whereas at $\Delta\varphi=\pi\pm 2\pi n$, the energy gain imparted to electrons in one region is negated in the second (here $n\in\mathbb{Z}$).

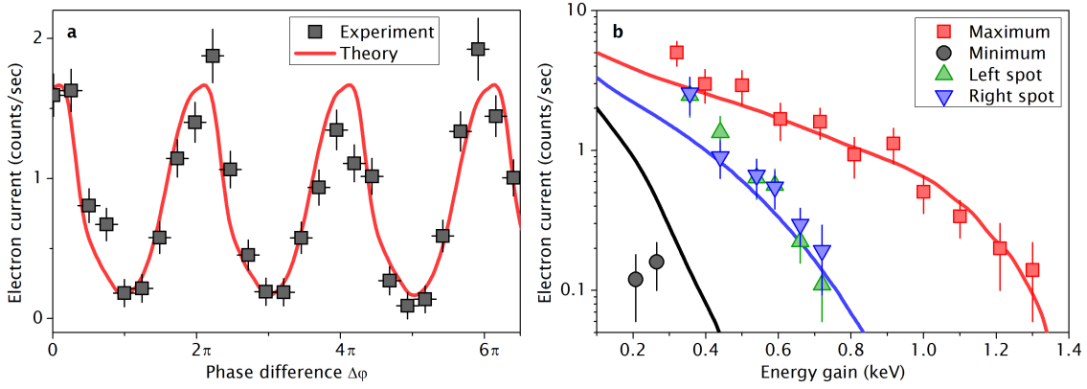

**Figure 2. Demonstration of phase-controlled two-stage acceleration. a.** The dependence of the accelerated electron current on the relative phase between the two driving optical pulses for a minimum energy gain of 400 eV (black) along with numerical results. **b.** The accelerated electron count rate as a function of the minimum energy gain for the relative phase for maximum energy gain (black squares), minimum energy gain (red circles) and for acceleration by each individual laser pulse (green and blue triangles) compared to the numerical results (lines). Acceleration in the first region can be doubled or cancelled in the second region depending on the relative phase of the two pulsed laser beams. However, we do not observe an exactly doubled energy gain of 1.5 keV due to deflection of electrons away from the surface of the structure, dephasing of electrons with significantly increasing velocities and the electronic noise of our detection system.

The crucial property of multi-stage acceleration is the increase of the total energy gain obtained by particles traversing several acceleration regions where each region operates close to the maximum acceleration gradient imposed by its damage threshold. Figure 2b shows the accelerated electron current as a function of the minimum energy gain for the relative phase differences $\Delta\varphi=\pm 2\pi n$ (red squares) and $\Delta\varphi=\pi\pm 2\pi n$ (black circles) between the two laser pulses. We further show the acceleration obtained in each individual interaction region. The peak laser fluence in each region is 0.13 J/cm$^2$, directly below the measured damage fluence of 0.14 J/cm$^2$. The maximum energy gain of 750 eV for electrons interacting with each section individually (blue and green triangles) is almost doubled to an energy gain of 1.3 keV when electrons interact with both pulses at the ideal phase difference. This result suggests that the electron energy gain grows linearly with the number of acceleration stages and emphasizes the importance of sub-optical cycle phase control in any DLA multi-stage device[23].

A DLA-based linear accelerator with a final energy of 1 MeV can be as small as 1 mm. However, this requires a nanofabricated source region, limiting the injected electron energy to tens of keV. Hence, the electron beam needs to be bunched and pre-accelerated at sub-relativistic energies. The former occurs due to the induced velocity distribution on the electron beam. The latter must take into account the dephasing of accelerated electrons: as sub-relativistic electrons are accelerated, their velocity increases significantly and soon phase slippage between the laser drive and the electron beam sets in[2,24]. Electrons injected at accelerating phases encounter decelerating phases after a length determined by the electron energy, laser wavelength, and field gradient. This must be compensated for by either frequency chirping the incident laser pulse or modifying the acceleration structure design, where the phase velocity of the accelerating mode has to match the particle velocity at all longitudinal coordinates. Since the former is limited by the spectral width of the laser pulse, here we use lithography-based structural chirping instead.

In our chirped structural design, the distance between grating teeth depends on the longitudinal coordinate $z$ as $\lambda_p=\lambda_{p0}+az$, where $\lambda_{p0}=620$ nm is the starting grating period and $a$ the chirp parameter. In Figure 3 we show the accelerated electron current as a function of $eU_s-E_{k0}$ for different values of $a$ but the same laser parameters, with $e$ the electron charge and $U_s$ the DC voltage applied to the spectrometer. The maximum energy gain is now limited by the incident field amplitude of $E_{max}=1.3$ GV/m and the interaction distance which in turn is given by the laser spot radius $w=29\pm3$ μm. Without any structural chirp ($a=0$), dephasing limits the maximum achievable energy gain to $\Delta E_k=0.80$ keV, in agreement with theory[24]. For $a=6.5*10^{-4}$, the maximum energy gain is more than tripled to $\Delta E_k=2.6$ keV, corresponding to an acceleration gradient of 69 MeV/m.

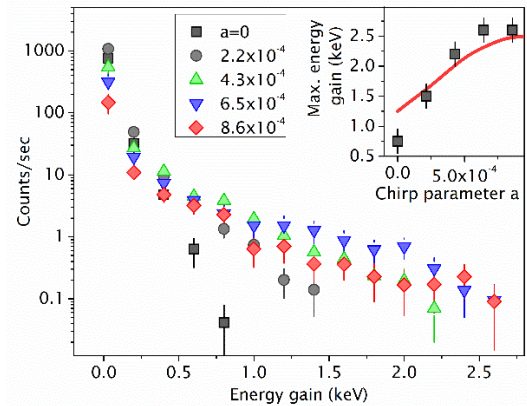

**Figure 3. Demonstration of sustained energy gain via introduction of a structure chirp.** Accelerated electron count rate as a function of the minimum energy gain ($eU_s-E_{k0}$) for different values of the chirp parameter $a$. Inset: The maximum energy gain as a function of $a$ (points) compared to numerical results (red curve).

In any accelerator of a certain length, active beam steering and collimation units are required[2]. In order to control the transverse dynamics of electrons traversing DLA-structures, focusing structures like that presented in Figure 4a must be employed. A deflecting mode is locally excited at a grating tilted by an angle with respect to the electron beam[12]. As the deflection angle $\varphi$ depends linearly on the grating tilt angle $\alpha$, the focusing effect can be reached by curving the grating teeth with an angle that depends on the distance from the center of design axis as $\alpha=\sigma x$, where $\sigma$ is a constant and $x$ is the distance from the design axis. This leads to parabolic grating teeth (see Figure 4a).

We demonstrate focusing by measurement of the dependence of the accelerated electron deflection angle $\varphi$ on the position of the electron beam center $x$ within the interaction region (see Figure 4b). We observe a linear dependence $\varphi=bx$, with the focal length $f=1/b$. Because the synchronous mode simultaneously deflects and accelerates in this structure[12], the focal distance differs for electrons with different energy gains (different colors in Figure 4d). We find that the effective focal lengths of electrons with minimum energy gains of 200, 400, and 600 eV are 500±20, 260±10 and 190±10 μm respectively for $\sigma=0.39$ μm$^{-1}$ and note that the corresponding focusing strength is sufficient for keeping the expected electron beam in multistage DLAs well-collimated.

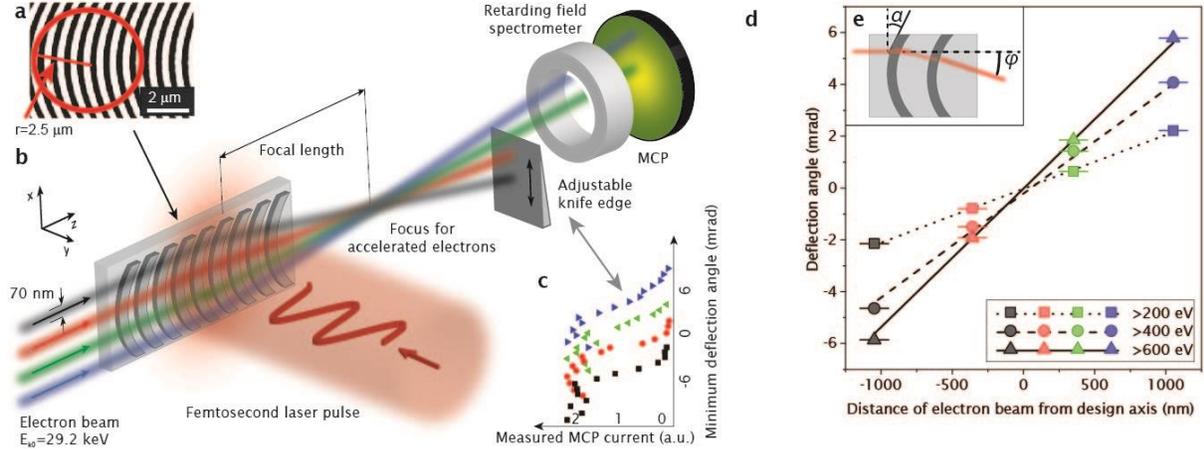

**Figure 4. Demonstration of a dielectric laser focusing structure. a.** A scanning electron microscope image of the dielectric laser focusing element, with parabolic grating teeth fabricated from Si. The radius of curvature at the vertex of the parabola, 2.5µm, is indicated. **b.** The characterization of focusing performance: accelerated electrons that traverse the lens above the parabolic vertex are deflected downwards and those that traverse below the vertex are deflected upwards. The grating curvature angle $\alpha$ and electron beam deflection angle $\varphi$ are shown in inset **e**. The spatial profile of the accelerated electrons as a function of $x$ is measured with a knife edge scan, sample results of which are shown in **c**. **d.** The position of the centroid of the accelerated spatial distribution at the location of the knife edge as a function of $x$, with linear fits for each energy setting. Due to the chromaticity of the lens, electrons that are accelerated strongly are also focused strongly, and thus the measured focal distance depends on the electron energy gain.

We have designed and experimentally demonstrated several critical components of a 2 µm fiber laser-based DLA. Note that for relativistic electrons ($\geq$ 5 MeV), high damage threshold material[25] (e.g. $CaF_2$) and identical laser parameters, we expect an average acceleration gradient in excess of 1.2 GeV/m and (for 5 MeV electrons) a focal length of 300 µm due to the more efficient coupling of laser light into the appropriate transverse and longitudinal modes. By using few cycle laser pulses, gradients of 4 GeV/m will be realized. These gradients and focal lengths outperform typical RF cavities and focusing elements (e.g. solenoids and permanent magnetic quadrupoles) by factors of 100 and 30 respectively, potentially leading to much smaller linear accelerators.

Further, much higher energy gains will be achieved through additional stages, the foundation of which we have demonstrated here. The principle questions that remain include the incorporation of an electron source appropriate for DLA operation[26-28], integration of the elements already demonstrated, and the determination of the radiative hardness of the dielectric structures. Closely tied to the source, the electron current that can be supported in such a structure depends on Coulomb repulsion[24], the incident laser profile, and the electron generated wakefields, requiring involved numerical investigation outside of the scope of this letter. However, the linear scaling of the average current with the repetition rate of the incident laser beam is promising due to the commercial availability of high repetition rate femtosecond (fiber) lasers.

This work thus paves the way towards realization of a potentially revolutionary species of linear accelerator: one whose acceleration gradient enables both a $cm^3$-sized 10 MeV medical proximity radiation device and the generation of GeV electrons within a university laboratory. Such devices may enable a new regime for modern accelerator physics and the many disciplines dependent on the charged particle beams produced therein, such as photon sources in all spectral ranges.

Acknowledgements

The authors acknowledge funding from ERC grant "Near Field Atto", the German BMBF Eurostars project "MIRANDUS" (E! 6698), and the Gordon and Betty Moore Foundation. Device fabrication was performed in the Stanford Nanofabrication Facility and the Stanford Nano Shared Facilities.

Author Contributions
J. McN and M. K. carried out the experiments, designed the samples tested, performed the simulations and created the figures in the manuscript. J. McN, M. K. and P.H. designed the experiments, interpreted the results, and wrote the manuscript. K. J. L., A. C., and H. D. prepared the silicon gratings. N.S, H. H., A. R., R. H., and I. H. developed and provided the lasers for the experiments. J. S.H, O.S. and R. L. B. supervised the work in Stanford and P. H. in Erlangen. J. McN and M. K. are co-first authors.